\documentclass[twocolumn,prl,aps,showpacs,preprintnumbers,amsmath,amssymb,superscriptaddress]{revtex4}
\usepackage{graphicx}
\begin{document}
\title{Two-Stage Formation Model and Helicity of Gold Nanowires}
\author{Yusuke Iguchi}
\affiliation{Department of Applied Physics, The University of Tokyo, 
Bunkyo-ku, Tokyo, 113-8656, Japan}

\author{Takeo Hoshi\footnote{Present address : Department of Applied 
Mathematics and Physics, Tottori University, Tottori 680-8550, Japan}}
\affiliation{Department of Applied Physics, The University of Tokyo, 
Bunkyo-ku, Tokyo, 113-8656, Japan}
\affiliation{Core Research for Evolutional Science and Technology, 
Japan Science and Technology Agency (CREST-JST), Japan}

\author{Takeo Fujiwara\footnote{Also at:
Center of Research and Development for Higher Education, 
The University of Tokyo, 
Bunkyo-ku, Tokyo, 113-8656, Japan}
}
\affiliation{Department of Applied Physics, The University of Tokyo, 
Bunkyo-ku, Tokyo, 113-8656, Japan}
\affiliation{Core Research for Evolutional Science and Technology, 
Japan Science and Technology Agency (CREST-JST), Japan}

%
\begin{abstract}
A model for formation of helical multishell gold nanowires is proposed 
and is confirmed  with the quantum mechanical 
molecular dynamics simulations.  
The model can explain the magic number of the helical gold nanowires 
in the multishell structure. 
The reconstruction from ideal non-helical to realistic helical nanowires consists 
of two stages: 
dissociations of atoms on the outermost shell from atoms on the inner shell 
and slip deformations of atom rows 
generating (111)-like structure on the outermost shell.
The elementary processes are governed by competition between energy loss and gain 
by s- and d-electrons together with the width of the d-band. 
The possibility for the helical nanowires of platinum, silver and copper 
is discussed. 
\end{abstract}
\date{\today}
\maketitle

Gold nanowires synthesized by electron beam technique 
have helical multishell structures.~\cite{kondo2000, oshima2003}
They are formed along the original [110] axis. 
The outermost shell is a (111)-like atomic sheet and 
 [110] atomic rows are helical around the nanowire axis.
The histogram of appearance of the diameters of nanowires 
has peaks of the numbers of atoms discretely 
at seven, eleven, thirteen, fourteen and so on.  
The difference of numbers of atoms between the outermost 
and  the next outermost shells is seven, called ``magic number'',
except cases of five and seven atoms on the outermost shell.
Platinum nanowires were also reported to have 
the same type of helicity.~\cite{oshima2002} 
Therefore, the helical multishell structure of nanowires might be 
generic not only for gold but also for a certain class of metals.
A structure of nanowires should be determined generally 
by competition between bulk and surface regions. 
A pioneering theoretical work~\cite{Gulseren1998}
predicted, with classical potentials of Al and Pb, that 
such nanowires can form various noncrystalline \lq exotic' structures
 including helical ones. 
There are several theoretical works on [110] helical multishell gold nanowires, 
based on quantum mechanical simulation.~\cite{tosatti2001,senger2004,yang2004}
Tossatti {\it et al.} reported, with the first principles calculations, 
that the tension of nanowires gives the minimum values  
when the number of atoms of the lateral atomic row on the outermost shell 
is seven and the nanowire is helical.~\cite{tosatti2001} 
They also showed that the tension does not have the minimum in model silver nanowires. 
Senger {\it et al.} achieved the first principles electronic structure calculations 
of nanowires with atoms from three to five on the lateral atomic row 
on the outermost shell, 
and showed that helical nanowires are not the configuration 
of the minimum energy but of the minimum tension.~\cite{senger2004}
In spite of these intensive studies,
no theories so far explain the formation process of the helicity and 
the magic number of helical nanowires of gold and platinum. 
In the present Letter, we propose a two-stage model of formation 
of multishell helical nanowires 
and explain the origins of the helicity and the magic number.
%

\begin{figure}
\begin{center}
\includegraphics[width=0.8\linewidth]{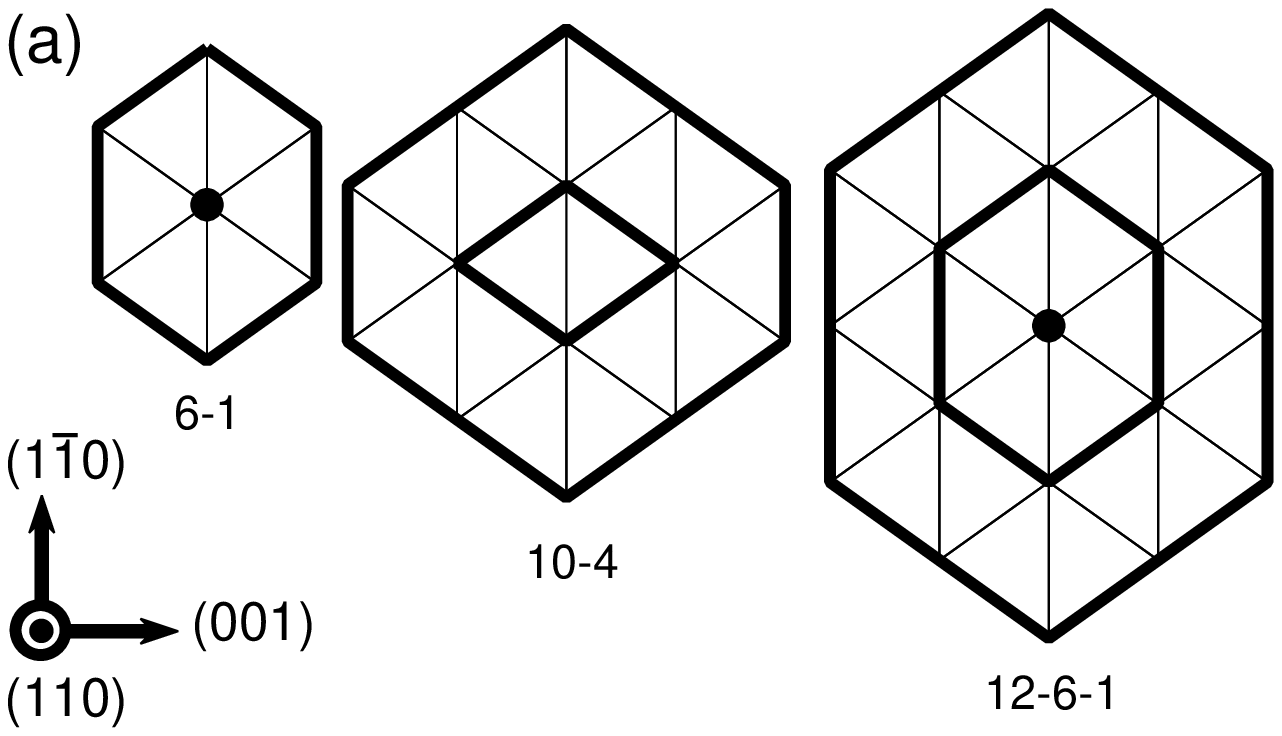}
\includegraphics[width=0.8\linewidth]{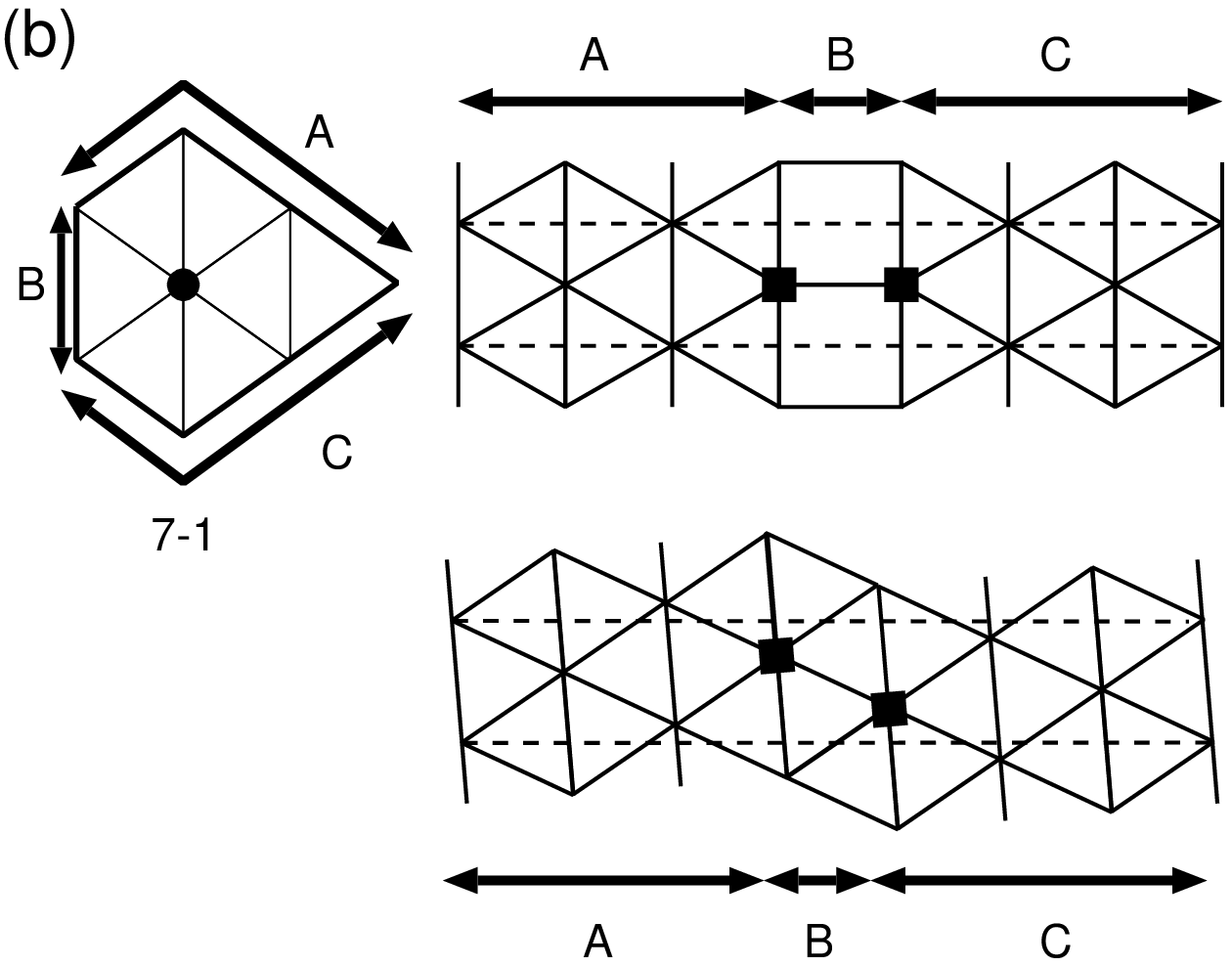}
\end{center}
\caption{
\label{fig:model}
(a) The ideal [110] fcc nanowire is intrinsically a non-helical multishell nanowire.
(b) Expanded a surface of the 7-1 multishell nanowire before (upper right) 
and after (lower right) surface reconstruction. 
Dashed lines connect the same atoms at the right and the left ends.
}
\end{figure}

%
In the ideal [110] fcc nanowire of stacking (110) sections, 
there are atom rows parallel to the [110] axis and 
we can find multishell structure with no helicity.
Figure~\ref{fig:model}(a) depicts 
three structures of the ideal fcc [110] nanowires, 6-1, 10-4, and 12-6,  
where the index stands for the number of atoms on the lateral atom row 
on the outermost shell, that on the next outermost shell and so on. 
The structure of these ideal model nanowires satisfies two conditions: 
(a) there is no acute angle on the surface because of 
diminishing surface tension, and
(b) there is no $(001)$ side longer than any (111) side 
since the surface energy of a (001) surface is higher than that of (111).
%

%
In the present model, the driving force for the helicity is 
the atom row slip. 
The helical structure is introduced when the outermost shell transforms 
into a folded (111)-like sheet. 
The transformation consists of two stages. 
Firstly the outermost shell should be dissociated 
from the inner shell to move freely and 
atom rows along [110] might be inserted into the outermost shell (Stage 1). 
Secondly, since a (111) sheet is energetically favorable on the outermost shell, 
an atom row slips  and (001) faces transform into (111) ones (Stage 2).
The outermost shell has six more ``bonds'' on the lateral row  
than the inner shell. 
See 6-1, 10-4 and 12-6-1 nanowires, shown by bold lines 
in Fig.~\ref{fig:model}(a). 
We add one another atom row on the outermost shell 
as in Fig.~\ref{fig:model}(b), 
then the outermost shell has seven more atoms 
on a lateral atom row than the inner shell and 
the outermost shell can have room for atom row slip. 
This is the origin of the ``magic number''.
When the number of atoms on the lateral row in the outermost shell is odd, 
the surface reconstruction brings the helicity to nanowires inevitably.  
Figure~\ref{fig:model}(b) shows the case of forming 7-1 helical nanowire where  
a slip of atom row transforms the surface region $B$, 
initially a (100)-like surface, into a (111)-like surface 
and a folded (111)-like surface brings the helicity.
Helical nanowires of 7-1, 11-4, 14-7-1 and 15-8-1 are 
experimentally observed.~\cite{kondo2000} 
%

\begin{figure}
\includegraphics[width=\linewidth]{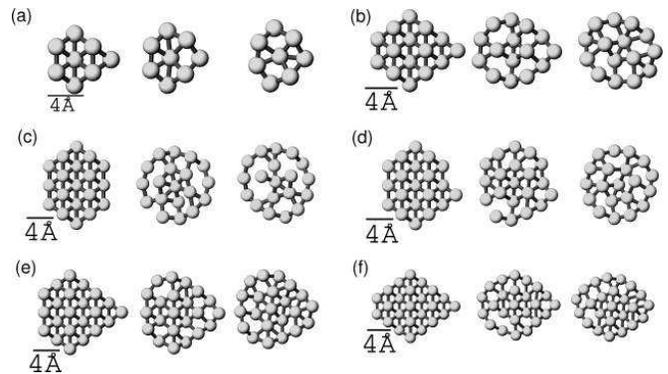}
\caption{
\label{fig:sections}
The sections of gold nanowires:
(a) 7-1 at the initial (left), 400 (middle) and 5,000~MD steps (right), 
(b) 11-4 at the initial (left), 400 (middle) and 6,000~MD steps (right),
(c) 12-6-1 at the initial (left), 7,500 (middle) and 13,000~MD steps (right),
(d) 13-6-1, at the initial (left), 750 (middle) and 9,000~MD steps (right),
(e) 14-7-1, at the initial (left), 400 (middle) and 7,000~MD steps (right),
and (f) 15-8-1 at the initial (left), 400 (middle) and 5,000~MD steps (right).
An atom row slips on the outermost shell and helicity is introduced.
Temperatures in the simulations are 600~K in (a), (b), (e), and (f) and 900~K 
in (c) and (d). 
}
\end{figure}

%
We performed, to verify the above model, the MD simulations of gold nanowires 
with the tight-binging (TB) Hamiltonian.~\cite{mehl1996,kirchhoff2001}
These TB parameters are prepared to represent electronic structures of bulk solids, 
surfaces, stacking faults and point defects, 
and have been tested in several other simulations 
for gold nanowires.~\cite{dasilva2001,dasilva2004,haftel2006}
We stack ideal (110) sections  of the fcc lattice, 
as an initial state, as depicted in Fig.~\ref{fig:sections}; 
nine layers in  the cases of 7-1 and 11-4, 
seven layers in the cases of 12-6-1 and 13-6-1,
and six layers in the cases of 14-7-1, and 15-8-1.
Then the numbers of atoms are from 76 to 156.
We adopt the boundary condition of fixing 
the center of gravity of both end layers of the nanowires,
and impose no external force. 
One MD step is 1~fs.
Figure \ref{fig:sections} shows the time evolution of typical sections 
in gold nanowires.
The outermost shell becomes a smooth surface and some atoms on it 
dissociate from atoms on the inner shell. 
The surface reconstruction is always observed, 
and no significant difference is observed 
at two different temperatures, 600~K and 900~K. 
In the case of 12-6-1 nanowire (Fig.~\ref{fig:sections}(c)), 
no surface reconstruction is observed at 600~K.  
%

\begin{figure}
\begin{minipage}{\linewidth}
\includegraphics[width=\linewidth]{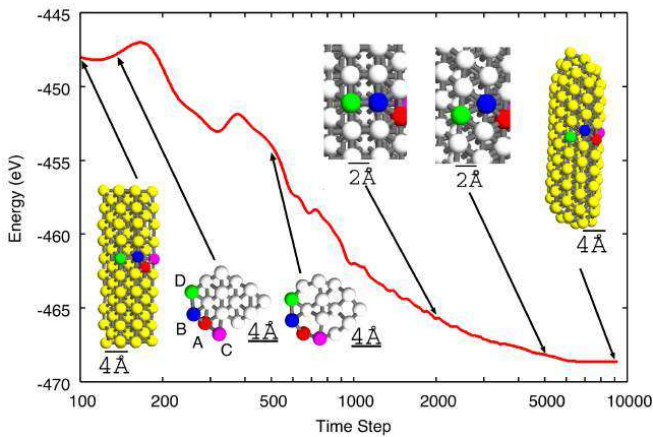}
\end{minipage}
\caption{
\label{fig:11-4}
Change of the total energy 
and the atomic configuration 
during the formation process of  the 11-4 nanowire at 600 K. 
Red, blue, purple, and green atoms are named A, B, C, and D respectively.
Within 500~MD steps, the atom A was disassociated from inner shell. 
Between 2,000 and 5,000~MD steps, the slip between the atom B and D 
brought the surface reconstruction, and (001) face becomes (111) face. 
}
\end{figure}

%
A typical time evolution of MD simulations is shown in Fig.~\ref{fig:11-4} 
for an 11-4 nanowire (Fig.~\ref{fig:sections}(b)) at 600 K. 
See also the supplementary movie.~\cite{fujiwaralab}
The initial state is a 10-4 nanowire 
with an additional atom row, and the total number of atoms is 143.
The total energy decreases almost monotonically after 1,000 MD steps.
The atom A in Fig.~\ref{fig:11-4}, which is on the (111) sheet 
and also its nearest neighbor atoms (e.g. B) on the (100) sheet, 
dissociates from the inner shell before 500 MD step (Stage 1).~\cite{dissociation}
Then the surface atoms can move independently from inner shell. 
From 2,000 to 5,000 MD steps,
(001) sheet reconstructs into hexagonal (111)-like surface
with a slip deformation,
and the helical structure appears (Stage 2). 
The inner shell rotates at the same time. 
These behaviors with two stages agree with the proposed model. 
It should be noticed that the time evolution in the MD simulation 
does not totally correspond to a real formation process of gold nanowires.  
But we believe that the real formation process has two different stages 
of short time process (Stage 1) and long time process (Stage 2). 
%

%
We are focusing on the change of the electronic structures  
to know why the atom A dissociates from the inner shell in Stage 1, 
and why the slip deformation appears around the atom B in Stage 2.
The local electronic density of states (LDOS) is shown 
in the local coordinate system in which 
the $x$-axis is the nanowire axis, $[110]$, 
the $y$-axis is along the lateral direction and 
the $z$-axis is perpendicular to the surface, so that 
the $xy$-orbital should be on the surface on the nanowire.
First we discuss the change of LDOS of the atoms A and C 
during the dissociation of the atom A (Stage 1).

\begin{figure}
\begin{center}
\includegraphics[width=\linewidth]{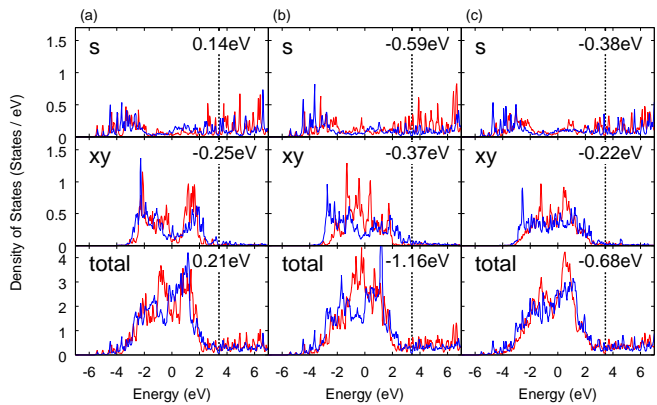}
\end{center}
\caption{
\label{fig:ldos}
Partial densities of states and local energies in orthogonalized basis.
Partial densities of states of (a) the atom A, (b) the atom C, and (c) the atom B.
Red and blue lines are at the initial state and at 500MD steps respectively in (a) and (b), 
and at 500MD steps and 5,000MD steps in (c).
Upper, middle, and lower figures are the partial densities of states 
of s-orbital, $xy$-orbital, and the total density of states respectively. 
The energy changes of the densities of states 
are written at the upper right of each figure.
Local coordinate system for each atom is written in text.
Notice that, in all cases, the asymmetry of LDOS of 
$xy$ orbital is enhanced.
}
\end{figure}

%
The dissociation of the atom A from the inner atoms 
changes the electronic structure of the atom A in Stage 1 as follows.  
The local energy of s-orbital increases
because of reduction of its coordination number 
(Fig.~\ref{fig:ldos}(a) top).
The local energy in $yz$-orbital of the atom A (not shown) increases
since the orbital extents  perpendicularly to the crystal surface 
and the nearest neighbor distance increases. 
The local energy in $xy$-orbital decreases, 
since the orbital can expand more to another (111) sheet 
through the atom C because of flattening two (111) sheets 
and the band width becomes wider (Fig.~\ref{fig:ldos}(a) middle).
Then the energy loss and gain of the d-orbitals almost cancel 
with each other. 
The net loss of the local energy of the atom A is mainly due to
energy loss of the s-orbital (Fig.~\ref{fig:ldos}(a) bottom).
%

%
Figure~\ref{fig:ldos} (b) shows LDOS of the atom C, 
the nearest neighbor of the atom A. 
There is relatively small energy loss of $yz$-orbital in Stage 1 
since the coordination number of the atom C does not change. 
The local energy of $xy$-orbital decreases (Fig.~\ref{fig:ldos}(b) middle).
Therefore, the local energy of the d-orbitals decreases on the atom C. 
This energy gain of d-orbitals can be attributed to 
the flattened surface structure around the atom C after 
the dissociation between the atom A and the inner shell.  
The energy of s-orbital of the atom C also decreases 
appreciably (Fig.~\ref{fig:ldos}(b) top) but 
it may be not associated primarily with the dissociation 
since the s-orbital does not always favor the flatter atomic configurations.   
The  dissociation between the atom A and the inner shell in Stage 1 
is governed by competition 
between the s-orbital energy loss of the atom A with reducing the coordination number  
and the energy gain of the d-orbitals due to the flattened surface.
The change in the energy of d-orbitals is almost proportional 
to the d-band width, 
and gold has wider d-band width than that of copper and silver.
Therefore, the bond dissociation of Stage 1 is preferable for gold 
but not so for silver and copper.
In fact, the MD simulation in copper does not show 
the bond dissociations of Stage 1 within 500~MD steps.

The electronic structure of the atom B changes 
as follows during the slip deformation (Stage 2). 
LDOS of the atom B is depicted 
in Fig.~\ref{fig:ldos} (c).  
The local energy of the s-orbital decreases 
since the coordination number of the atom B increases from seven to eight 
(Fig.~\ref{fig:ldos} (c) top).
The local energy in the $xy$-orbital of the atom B also decreases 
since the surface transforms to (111)-like and is flattened 
(Fig.~\ref{fig:ldos} (c) middle).
Here, the slip deformation is essential since it widens the area of the 
(111)-like triangular surface and 
LDOS of $xy$ orbital transfers 
its weight from the anti-bonding region (the high energy region) to 
the bonding region (the low energy region) 
(Figs.~\ref{fig:ldos} (a), (b) and (c) middle). 
Therefore, both orbitals s and $xy$ of the atom B result 
in decrease of the local energy.
The accomplishment of slip deformation is, however, 
depends on the details of surrounding atoms. 
The atom A connecting with the atom B 
can move relatively freely from the inner shell 
since the bondings with the inner shell are dissociated after Stage 1.
Therefore, the atom B slips easily and creates the helicity  
since the atom B can trail other atoms without dissociating their bonds.
Higher temperatures were required for reconstruction of (001) surface 
in  copper nanowires 
since the atom A in copper does not dissociate from the inner shell.
%

%
The two-stage formation model holds also in 12-6-1 and 13-6-1 nanowires. 
Figures~\ref{fig:sections}(c) and (d) show how inserted atom row would behave. 
See supplementary movie.~\cite{fujiwaralab}
Both  nanowires show surface reconstruction,
and the number of atoms in the outermost shell is 13.
Therefore, the inserted atoms are supplied possibly from outer and inner shells 
and positions of supplied atoms is not seriously crucial to 
the surface reconstruction, or the atom row slip.  
In all cases, 
the bond dissociation between the outermost atoms and the inner shell
starts near a cross edge of two ideal faces, 
since the energy gain is maximized by flattening a surface on the edges. 
The ratio of the energy gain to the total energy 
due to the dissociation 
becomes smaller in thicker nanowires and 
then the thicker nanowires do not transform.  
MD simulation for 11-4 and 12-6-1 nanowires in copper 
does not show the bond dissociation from the inner shell 
of Stage 1 at 600~K within 500~MD steps, since copper has narrower d-band width.  
At 900~K the atom row slips on the (001) sheet in copper nanowires, 
since the atom have dissociated from the inner shell at that temperature.
The present analysis shows that the mechanism in both
Stage 1 and Stage 2 is governed by the d-band width and
the helical nanowires appear only among metals with a
wider d-band. 
The d-band width in platinum and gold is
commonly wider than that in lighter elements, Ag and Cu,
and the present theory explains why platinum nanowire
can be also formed with helicity.
Finally, we point out that
the present mechanism is inherent not only
with nanowires but also with the bulk surface, 
e.g. the Au (001) surface reconstruction to Au (001)-hex, 
hexagonal close-packed one.~\cite{binnig1984, abernathy1992} 
This also implies a reason
why (001) surface reconstruction is observed in metals
with 5d orbitals such as Au, Pt, and Ir, but not observed
in metals with the outer 3d or 4d shell such as Ag and Cu.
%


In summary, 
we have proposed the two-stage model of
formation of gold multishell helical nanowires and its 
\lq magic number'. 
The model is confirmed by molecular dynamics 
simulation with electronic structure and
its mechanism is explained as  
a nanometer-scale size effect 
dominated by d-orbitals on nanowire surface.
Helicity is introduced by the surface reconstruction or the
slip of an atom row on the (001) sheet, because the
triangular (111)-like sheet is  more preferable  for d-orbitals
extending over the surface. 
%



\begin{thebibliography}{15}
\expandafter\ifx\csname natexlab\endcsname\relax\def\natexlab#1{#1}\fi
\expandafter\ifx\csname bibnamefont\endcsname\relax
  \def\bibnamefont#1{#1}\fi
\expandafter\ifx\csname bibfnamefont\endcsname\relax
  \def\bibfnamefont#1{#1}\fi
\expandafter\ifx\csname citenamefont\endcsname\relax
  \def\citenamefont#1{#1}\fi
\expandafter\ifx\csname url\endcsname\relax
  \def\url#1{\texttt{#1}}\fi
\expandafter\ifx\csname urlprefix\endcsname\relax\def\urlprefix{URL }\fi
\providecommand{\bibinfo}[2]{#2}
\providecommand{\eprint}[2][]{\url{#2}}

\bibitem[{\citenamefont{Kondo and Takayanagi}(2000)}]{kondo2000}
\bibinfo{author}{\bibfnamefont{Y.}~\bibnamefont{Kondo}} \bibnamefont{and}
  \bibinfo{author}{\bibfnamefont{K.}~\bibnamefont{Takayanagi}},
  \bibinfo{journal}{Science} \textbf{\bibinfo{volume}{289}},
  \bibinfo{pages}{606} (\bibinfo{year}{2000}).

\bibitem[{\citenamefont{Oshima et~al.}(2003)\citenamefont{Oshima, Onga, and
  Takayanagi}}]{oshima2003}
\bibinfo{author}{\bibfnamefont{Y.}~\bibnamefont{Oshima}},
  \bibinfo{author}{\bibfnamefont{A.}~\bibnamefont{Onga}}, \bibnamefont{and}
  \bibinfo{author}{\bibfnamefont{K.}~\bibnamefont{Takayanagi}},
  \bibinfo{journal}{Phys. Rev. Lett.} \textbf{\bibinfo{volume}{91}},
  \bibinfo{pages}{205503} (\bibinfo{year}{2003}).

\bibitem[{\citenamefont{Oshima et~al.}(2002)\citenamefont{Oshima, Koizumi,
  Mouri, Hirayama, Takayanagi, and Kondo}}]{oshima2002}
\bibinfo{author}{\bibfnamefont{Y.}~\bibnamefont{Oshima}},
  \bibinfo{author}{\bibfnamefont{H.}~\bibnamefont{Koizumi}},
  \bibinfo{author}{\bibfnamefont{K.}~\bibnamefont{Mouri}},
  \bibinfo{author}{\bibfnamefont{H.}~\bibnamefont{Hirayama}},
  \bibinfo{author}{\bibfnamefont{K.}~\bibnamefont{Takayanagi}},
  \bibnamefont{and} \bibinfo{author}{\bibfnamefont{Y.}~\bibnamefont{Kondo}},
  \bibinfo{journal}{Phys. Rev. B} \textbf{\bibinfo{volume}{65}},
  \bibinfo{pages}{121401(R)} (\bibinfo{year}{2002}).


\bibitem[{\citenamefont{Gulseren et~al.}(1998)}]{Gulseren1998}
\bibinfo{author}{\bibfnamefont{O.}~\bibnamefont{G\"ulseren}},
\bibinfo{author}{\bibfnamefont{F.}~\bibnamefont{Ercolessi}},
\bibinfo{author}{\bibfnamefont{E.}~\bibnamefont{Tosatti}},
\bibinfo{journal}{Phys. Rev. Lett.} \textbf{\bibinfo{volume}{80}}, \bibinfo{pages}{3775}
(\bibinfo{year}{1998}).




\bibitem[{\citenamefont{Tosatti et~al.}(2001)\citenamefont{Tosatti, Prestipino,
  Kostlmeier, {Dal Corso}, and {Di Tolla}}}]{tosatti2001}
\bibinfo{author}{\bibfnamefont{E.}~\bibnamefont{Tosatti}},
  \bibinfo{author}{\bibfnamefont{S.}~\bibnamefont{Prestipino}},
  \bibinfo{author}{\bibfnamefont{S.}~\bibnamefont{Kostlmeier}},
  \bibinfo{author}{\bibfnamefont{A.}~\bibnamefont{{Dal Corso}}},
  \bibnamefont{and} \bibinfo{author}{\bibfnamefont{F.~D.} \bibnamefont{{Di
  Tolla}}}, \bibinfo{journal}{Science} \textbf{\bibinfo{volume}{291}},
  \bibinfo{pages}{288} (\bibinfo{year}{2001}).

\bibitem[{\citenamefont{Senger et~al.}(2004)\citenamefont{Senger, Dag, and
  Ciraci}}]{senger2004}
\bibinfo{author}{\bibfnamefont{R.~T.} \bibnamefont{Senger}},
  \bibinfo{author}{\bibfnamefont{S.}~\bibnamefont{Dag}}, \bibnamefont{and}
  \bibinfo{author}{\bibfnamefont{S.}~\bibnamefont{Ciraci}},
  \bibinfo{journal}{Phys. Rev. Lett.} \textbf{\bibinfo{volume}{93}},
  \bibinfo{pages}{196807} (\bibinfo{year}{2004}).

\bibitem[{\citenamefont{Yang}(2004)}]{yang2004}
\bibinfo{author}{\bibfnamefont{C.-K.} \bibnamefont{Yang}},
  \bibinfo{journal}{Appl. Phys. Lett.} \textbf{\bibinfo{volume}{85}},
  \bibinfo{pages}{2923} (\bibinfo{year}{2004}).


\bibitem[{\citenamefont{Mehl and Papaconstantopoulos}(1996)}]{mehl1996}
\bibinfo{author}{\bibfnamefont{M.~J.} \bibnamefont{Mehl}} \bibnamefont{and}
  \bibinfo{author}{\bibfnamefont{D.~A.} \bibnamefont{Papaconstantopoulos}},
  \bibinfo{journal}{Phys. Rev. B} \textbf{\bibinfo{volume}{54}},
  \bibinfo{pages}{4519} (\bibinfo{year}{1996}).

\bibitem[{\citenamefont{Kirchhoff et~al.}(2001)\citenamefont{Kirchhoff, Mehl,
  Papanicolaou, Papaconstantopoulos, and Khan}}]{kirchhoff2001}
\bibinfo{author}{\bibfnamefont{F.}~\bibnamefont{Kirchhoff}},
  \bibinfo{author}{\bibfnamefont{M.~J.} \bibnamefont{Mehl}},
  \bibinfo{author}{\bibfnamefont{N.~I.} \bibnamefont{Papanicolaou}},
  \bibinfo{author}{\bibfnamefont{D.~A.} \bibnamefont{Papaconstantopoulos}},
  \bibnamefont{and} \bibinfo{author}{\bibfnamefont{F.~S.} \bibnamefont{Khan}},
  \bibinfo{journal}{Phys. Rev. B} \textbf{\bibinfo{volume}{63}},
  \bibinfo{pages}{195101} (\bibinfo{year}{2001}).

\bibitem[{\citenamefont{Haftel and Gall}(2006)}]{haftel2006}
\bibinfo{author}{\bibfnamefont{M.~I.} \bibnamefont{Haftel}} \bibnamefont{and}
  \bibinfo{author}{\bibfnamefont{K.}~\bibnamefont{Gall}},
  \bibinfo{journal}{Phys. Rev. B} \textbf{\bibinfo{volume}{74}},
  \bibinfo{pages}{035420} (\bibinfo{year}{2006}). 

\bibitem[{\citenamefont{da~Silva et~al.}(2001)\citenamefont{da~Silva, da~Silva,
  and Fazzio}}]{dasilva2001}
\bibinfo{author}{\bibfnamefont{E.~Z.} \bibnamefont{da~Silva}},
  \bibinfo{author}{\bibfnamefont{A.~J.~R.} \bibnamefont{da~Silva}},
  \bibnamefont{and} \bibinfo{author}{\bibfnamefont{A.}~\bibnamefont{Fazzio}},
  \bibinfo{journal}{Phys. Rev. Lett.} \textbf{\bibinfo{volume}{87}},
  \bibinfo{pages}{256102} (\bibinfo{year}{2001}).

\bibitem[{\citenamefont{da~Silva et~al.}(2004)\citenamefont{da~Silva, Novaes,
  da~Silva, and Fazzio}}]{dasilva2004}
\bibinfo{author}{\bibfnamefont{E.~Z.} \bibnamefont{da~Silva}},
  \bibinfo{author}{\bibfnamefont{F.~D.} \bibnamefont{Novaes}},
  \bibinfo{author}{\bibfnamefont{A.~J.~R.} \bibnamefont{da~Silva}},
  \bibnamefont{and} \bibinfo{author}{\bibfnamefont{A.}~\bibnamefont{Fazzio}},
  \bibinfo{journal}{Phys. Rev. B} \textbf{\bibinfo{volume}{69}},
  \bibinfo{pages}{115411} (\bibinfo{year}{2004}).

\bibitem[{fuj()}]{fujiwaralab}
\bibinfo{note}{See EPAPS Document No.[number will be inserted by publisher]
for a movie of MD simulation of 11-4 nanowire at 600 K and 
for that of MD simulation of typical sections of 12-6-1 and 13-6-1 nanowires at 
900 K. 
This document is accessible through a direct link in the online article's 
HTML reference or the EPAPS homepage (http://www.aip.org/pubservs/epaps.html).}

\bibitem[{diss()}]{dissociation}]
\bibinfo{note}{Dissociation was ascertained in the present work in two ways; 
(1)~the interatomic distance increases by more than 20~\% and 
(2)~the peak height of the crystal orbital Hamiltonian population (COHP) 
decreases down to $1/5$ of that of the initial state. 
The COHP is defined in: R. Dronskowski and P. E. Bl\"ochl, J. Phys. Chem. 
{97}, 8617 (1993).} 

\bibitem[{\citenamefont{Binnig et~al.}(1984)\citenamefont{Binnig, Rohrer,
  Gerber, and Stall}}]{binnig1984}
\bibinfo{author}{\bibfnamefont{G.}~\bibnamefont{Binnig}},
  \bibinfo{author}{\bibfnamefont{H.}~\bibnamefont{Rohrer}},
  \bibinfo{author}{\bibfnamefont{C.}~\bibnamefont{Gerber}}, \bibnamefont{and}
  \bibinfo{author}{\bibfnamefont{E.}~\bibnamefont{Stall}},
  \bibinfo{journal}{Surf. Sci.} \textbf{\bibinfo{volume}{144}},
  \bibinfo{pages}{321} (\bibinfo{year}{1984}).

\bibitem[{\citenamefont{Abernathy et~al.}(1992)\citenamefont{Abernathy,
  Mochrie, Zehner, Gr\"ubel, and Gibbs}}]{abernathy1992}
\bibinfo{author}{\bibfnamefont{D.~L.} \bibnamefont{Abernathy}},
  \bibinfo{author}{\bibfnamefont{S.~G.~J.}~\bibnamefont{Mochrie}},
  \bibinfo{author}{\bibfnamefont{D.~M.} \bibnamefont{Zehner}},
  \bibinfo{author}{\bibfnamefont{G.}~\bibnamefont{Gr\"ubel}}, \bibnamefont{and}
  \bibinfo{author}{\bibfnamefont{D.}~\bibnamefont{Gibbs}},
  \bibinfo{journal}{Phys. Rev. B} \textbf{\bibinfo{volume}{45}},
  \bibinfo{pages}{9272} (\bibinfo{year}{1992}).



\end{thebibliography}
\end{document}